\documentclass[aps,pra,twocolumn,floatfix,superscriptaddress]{revtex4-2}

\usepackage{amsmath}
\usepackage{amssymb}
\usepackage{subfigure}
\usepackage{graphicx}

\begin{document}

\title{Structures and proximity effects of inhomogeneous population-imbalanced Fermi gases with pairing interactions}

\author{Bishal Parajuli}
\affiliation{Department of Physics, California Polytechnic State University, San Luis Obispo, California 93407, USA.}

\author{Devin J. Gagnon}
\affiliation{Department of Physics, California Polytechnic State University, San Luis Obispo, California 93407, USA.}

\author{Chih-Chun Chien}
\email{cchien5@ucmerced.edu}
\affiliation{Department of Physics, University of California, Merced, California 95343, USA.}

\begin{abstract}
By introducing spatially varying profiles of pairing interaction or spin polarization to quasi one-dimensional two-component atomic Fermi gases confined in box potentials, we analyze the ground state structures and properties when multiple phases coexist in real space by implementing the Bogoliubov--de~Gennes equation suitable for describing inhomogeneous fermion systems. While the BCS, Fulde--Ferrell--Larkin--Ovchinnikov (FFLO), and normal phases occupy different regions on the phase diagram when the parameters are uniform, a spatial change of pairing strength or spin polarization can drive the system from the FFLO phase to a normal gas or from a BCS superfluid to the FFLO phase in real space. The FFLO phase exhibits its signature modulating order parameter at the FFLO momentum due to population imbalance, and the pair correlation penetrates the polarized normal phase and exhibits proximity effects. Meanwhile, the BCS phase tends to repel population imbalance and maintain a plateau of pairing. Interestingly, a buffer FFLO phase emerges when the spatial change attempts to join the BCS and normal phase in the presence of spin polarization.
By analyzing the pairing correlations, interfacial properties, and momentum-space spectra of the inhomogeneous structures, relevant length- and momentum- scales and their interplay are characterized.
We also briefly discuss implications of inhomogeneous multi-phase atomic Fermi gases with population imbalance. 
\end{abstract}

\maketitle
\section{Introduction}

Conventional BCS Fermi superfluid with singlet pairing is resistant to spin polarization~\cite{Parks69,Tinkham}, making the search for a spin-polarized Fermi superfluid an interesting but challenging task. Among many possible phases, the Fulde--Ferrell--Larkin--Ovchinnikov (FFLO) phase~\cite{Fulde1964,Larkin1965} has been an intriguing option due to its modulating order parameter, which may accommodate population imbalance in real space.
While the BCS Cooper pairs have zero center-of-mass momentum, 
the FFLO features finite-momentum pairing due to the mismatched Fermi surfaces.
Recent experiments in layered superconductors and engineered superlattices have reported spectroscopic features consistent with finite-momentum pairing~\cite{Cho2025,Lin2025} while theoretical work has proposed the FFLO phases in low-dimensional cold-atom systems (see Ref.~\cite{Kinnunen2018RPP} for a review), spin-orbit-coupled Fermi gases~\cite{SeoZhangTewari2013, Zheng2014}, and flat-band materials~\cite{Sun2025}.

More recently, atomic Fermi gases have emerged as a powerful and versatile platform for exploring many-body physics~\cite{Pethick-BEC,Ueda2010,Zwerger2012}. Early cold-atom experiments have realized population-imbalanced Fermi gases with pairing interactions~\cite{Zwierlein2006,Partridge2006}. Comprehensive reviews of population-imbalanced Fermi gases and their rich phase structure can be found in Refs.~\cite{Giorgini2008,ChevyMora2010,Radzihovsky_2010}. Experimental signatures consistent with the FFLO phase have also been reported in quasi one-dimensional atomic Fermi gases~\cite{Liao2010,Revelle2016}. In parallel, momentum-resolved probes, including radio-frequency spectroscopy, Bragg spectroscopy, and noise-correlation measurements, have been proposed and partially implemented to address this challenge~\cite{Altman2004,Greiner2005,Gupta2003,Stewart2008}.

Meanwhile, the Bogoliubov--de Gennes (BdG) formalism~\cite{deGennes_book,Zhu-book} provides a manageable approach for investigating inhomogeneous Fermi gases with pairing interactions and spin polarization,
including the BCS, FFLO, and normal phases~\cite{Bakhtiari2008,SeoZhangTewari2013,KawamuraOhashi2022,VitaliZhang2022}.
The BdG approach is capable of visualizing spatial textures of the order parameter as well as the density, identifying proximity-induced structures and correlations, and connecting to experimentally measurable quantities in finite systems. The BdG approach has been applied to other topics in cold atoms, such as Bose-Fermi mixtures~\cite{Parajuli2023} and local potentials~\cite{PhysRevA.83.061604,Chien2024,Chien2025}.
Going beyond homogeneous systems, proximity effects arise when a parameter becomes inhomogeneous, leading to two sides of the system with different phases with an interface separating them.
A common example is the superconductor-normal-metal heterostructures~\cite{deGennes1969ProximityEffects,Falk1963,Silvert1964,Yamazaki2007}, where the penetration of pairing correlations into the normal region can lead to interesting behavior without genuine superconductivity there.
More complex situations of proximity effects involve the  superconductor-ferromagnet heterostructures featuring  oscillatory Cooper-pair amplitude~\cite{Buzdin2005,Bergeret2005}. Experimental works on the oscillatory pairing and interfacial properties have been reported in
Refs.~\cite{Ryazanov2001,Kontos2002}.
More recent developments in superconducting spintronics also illustrate how magnetic inhomogeneity reshapes pairing correlations and enables long-range spin-polarized supercurrents~\cite{LinderRobinson2015,Eschrig2015,Robinson2010Science,Banerjee2014}. Moreover, inhomogeneous phenomena have been explored in atomic Fermi gases with spatially inhomogeneous pairing interactions~\cite{Chien12,Piselli2018}.





On the other hand, inhomogeneous effects related to proximity effects have been studied by a different framework known as the spatial quench, where a spatial drop of a parameter drives a system across a critical point in real space. For the transverse field Ising model, Bose Hubbard model, and spinor Bose-Einstein condensate (BEC), the spatial Kibble-Zurek mechanism leads to scaling behavior across the interface (see Ref.~\cite{Dziarmaga2010} for a review). However, the order parameter of BCS superfluid exhibits non-analytic behavior and results in different correlations across the interface between the BCS and normal phases~\cite{Parajuli2023a}.

By introducing spatial quench to population-imbalanced Fermi gases, we will present proximity effects induced by inhomogeneous pairing interaction or spin polarization in atomic Fermi gases confined in box potentials. The box potentials provide a convenient platform for analyzing bulk properties of cold atoms~\cite{Gaunt2013,Mukherjee2017} and avoids trap-induced inhomogeneity, such as shell structures in harmonic traps when the potential depth changes.
There have been thermodynamic and spectroscopic measurements using box potentials~\cite{Yan2019,Ville2018}. As will be shown, the introduction of inhomogeneous parameters in box potentials joins different phases in real space and allows us to explore proximity effects without distraction from a varying background potential.

Different from previous work on proximity effects with equal population~\cite{Parajuli2023a}, the BdG formalism in this work will include different chemical potentials to account for population imbalance. We first present the phase diagram of attractive Fermi gases with homogeneous parameters to locate where the BCS, FFLO, and normal phases reside. By having different parameters in real space, we will present the structures of FFLO-normal, FFLO-BCS, and BCS-normal mixtures and analyze the pairing correlations in different regions. Importantly, the FFLO pair correlations with its signature modulations are shown to penetrate the normal region. Furthermore, we will show evidence of a buffer of the FFLO phase when the BCS and normal phases are joined together in the presence of spin polarization. We will also present the spin-resolved densities to confirm that the BCS phase resists population imbalance in its bulk while the FFLO and normal phases can accommodate spin polarization.

We remark that even when the pairing interaction and spin polarization are uniform, a highly polarized Fermi gas can phase separate into a BCS region and a polarized normal region when the polarization exceeds the Clogston-Chandrasekhar limit~\cite{PhysRevLett.9.266,10.1063/1.1777362}. Experimental results of population-imbalanced atomic Fermi gases with tunable pairing interactions~\cite{Zwierlein2006,Partridge2006} favor such a structure in three dimensions. In contrast, the multi-phase structures discussed here are due to inhomogeneous pairing interaction or spin polarization. Hence, the bulk of each phase is more stable and the interfacial and proximity effects are more prominent. Moreover, Fermi gases with strong pairing interaction on the BEC side of the BCS-BEC crossover can accommodate population imbalance in momentum space~\cite{Chien2007}, thereby showing a homogeneous profile. Here we focus on Fermi gases with intermediate pairing interaction in order to examine the interplay between the FFLO, BCS, and normal phases in the presence of population imbalance.

The rest of the paper is organized as follows. Sec.~\ref{Sec:Theory} outlines the theoretical treatment of two-component atomic Fermi gases with attractive interactions and population imbalance in one-dimensional box potentials via the BdG formalism. The length and momentum scales associated with the BCS, FFLO, and polarized normal phases are also discussed. Sec.~\ref{Sec:Results} presents the phase diagram of quasi one dimensional population-imbalanced attractive Fermi gases with uniform parameters and then shows selected configurations generated by inhomogeneous profiles of the pairing strength or spin polarization. Sec.~\ref{Sec:Implications} discusses possible experimental and theoretical implications. We conclude our work in Sec.~\ref{Sec:Conclusion}. The Appendix presents results for the FFLO phase with uniform parameters and boundary effects due to box potentials.

\section{Theoretical background}\label{Sec:Theory}
\subsection{Two-component attractive fermions in a box}
We consider a quasi one-dimensional system of two-component fermions labeled by spin indices $\sigma = \uparrow, \downarrow$ with tunable attractive contact interactions confined in a box. In the BCS mean-field approximation, the mean-field Hamiltonian is given by~\cite{Fetter_book,Pethick-BEC}.
\begin{align}
\mathcal{H}_{\text{MF}} = & \int dx\, (\sum_{\sigma}\psi_\sigma^\dagger(x)\, h_\sigma(x)\, \psi_\sigma(x) + \Delta(x)\, \psi_\uparrow^\dagger(x)\, \psi_\downarrow^\dagger(x) \nonumber \\
& + \text{H.c.} ).
\end{align}
Here the order parameter is 
\begin{equation} \label{Eq:Delta0}
\Delta(x) = -g\, \langle \psi_\downarrow(x) \psi_\uparrow(x) \rangle
\end{equation} 
with $g<0$ being the coupling constant for attractive interactions, and 
the single-particle Hamiltonian is
\begin{equation}
h_\sigma(x) = -\frac{\hbar^2}{2m} \frac{d^2}{dx^2} + V_{\text{ext}}(x) - \mu_\sigma.
\end{equation}
We introduce spin-dependent chemical potentials $\mu_\uparrow = \mu + h$ and $\mu_\downarrow = \mu - h$ and use $V_{ext}$ to model the box potential. 

Introducing the Nambu spinor
\begin{equation}
\Psi(x) = 
\begin{pmatrix}
\psi_\uparrow(x) \\
\psi_\downarrow(x) \\
\psi_\uparrow^\dagger(x) \\
\psi_\downarrow^\dagger(x)
\end{pmatrix},
\end{equation}
the Hamiltonian can be written as
\begin{equation}
\mathcal{H}_{\text{MF}} = \tfrac{1}{2}\int dx\, \Psi^\dagger(x)\, \mathcal{H}_{\text{BdG}}(x)\, \Psi(x)
\end{equation}
after dropping constant terms. Here
\begin{equation}
\mathcal{H}_{\text{BdG}}(x) =
\begin{pmatrix}
h_\uparrow & 0 & 0 & \Delta(x) \\
0 & h_\downarrow & -\Delta(x) & 0 \\
0 & -\Delta^*(x) & -h_\uparrow & 0 \\
\Delta^*(x) & 0 & 0 & -h_\downarrow
\end{pmatrix}.
\end{equation}

\subsection{Bogoliubov Transformation and Diagonalization}
The BdG Hamiltonian can be diagonalized via the Bogoliubov transformation:
\begin{align}
\psi_\uparrow(x) &= \sum_n \left[ u_\uparrow^{n1}(x)\,\gamma_{n1} - v_\uparrow^{n2*}(x)\,\gamma_{n2}^\dagger \right], \\
\psi_\downarrow(x) &= \sum_n \left[ u_\downarrow^{n2}(x)\,\gamma_{n2} + v_\downarrow^{n1*}(x)\,\gamma_{n1}^\dagger \right],
\end{align}
where $\gamma_{n1}$ and $\gamma_{n2}$ are quasiparticle annihilation operators.
The BdG Hamiltonian has a particle–hole symmetry~\cite{deGennes_book,Zhu-book} since  each eigenstate with $E_{n1}>0$ has a counterpart with $E_{n2}=-E_{n1}$. Their wavefunctions are related by
\begin{equation}
\begin{pmatrix}
u_\downarrow^{n2}(x) \\
v_\uparrow^{n2}(x)
\end{pmatrix}
=
\begin{pmatrix}
v_\downarrow^{n1*}(x) \\
- u_\uparrow^{n1}(x)
\end{pmatrix}.
\end{equation}
The corresponding quasiparticle operators are related by $\gamma_{n2}^\dagger \leftrightarrow \gamma_{n1}$. 
Utilizing the symmetry, we only need to focus on the positive-energy states and simplify the notation with $\gamma_n$ annihilating those positive-energy states and $\sum_n^\prime$ only summing over them.

The diagonalized mean-field Hamiltonian is
\begin{equation}
\mathcal{H}_{\text{MF}} = E_g + {\sum_{n}}' E_n\, \gamma_n^\dagger \gamma_n.
\end{equation} 
Here $E_g$ denotes the ground-state energy,
and the quasiparticle wavefunctions are determined by
\begin{equation}
\begin{aligned}
&
\mathcal{H}_{BdG}
\begin{pmatrix}
u_\uparrow^{n}(x) \\
u_\downarrow^{n}(x) \\
v_\uparrow^{n}(x) \\
v_\downarrow^{n}(x)
\end{pmatrix}
=
E_{n}
\begin{pmatrix}
u_\uparrow^{n}(x) \\
u_\downarrow^{n}(x) \\
v_\uparrow^{n}(x) \\
v_\downarrow^{n}(x)
\end{pmatrix}.
\end{aligned}
\label{eq:BdG-matrix}
\end{equation}
The order parameter given by Eq.~\eqref{Eq:Delta0} becomes
\begin{equation}\label{Eq:BdG_Delta}
\Delta(x) = -\frac{g}{2} {\sum_n}' \left[ 
u_\uparrow^n(x) \, v_\downarrow^{n*}(x) 
+ u_\downarrow^n(x) \, v_\uparrow^{n*}(x) 
\right].
\end{equation}
Moreover, the spin-resolved densities are given by
\begin{align}
\rho_\sigma(x) &= {\sum_n}' |v_\sigma^n(x)|^2. 
\end{align}

In practice, the Bogoliubov--de Gennes equations are solved self-consistently using an iterative scheme. One begins with an initial guess for the order parameter $\Delta(x)$, typically chosen as a uniform value or a simple spatial profile. Using this trial order parameter, the BdG Hamiltonian is diagonalized to obtain the quasiparticle wave functions and eigenenergies. These solutions are then used to update the order parameter via the gap equation \eqref{Eq:BdG_Delta}. 
The newly obtained $\tilde \Delta(x)$ is then used in the next round of iteration for the BdG equation. The procedure is repeated until convergence is reached, typically when the relative change in the gap function falls below a prescribed tolerance $\epsilon=\int|\tilde \Delta^{t+1}-\tilde \Delta^t|dx'$. In our numerical calculations, we use $\epsilon<10^{-5}$. This iterative approach guarantees that the final solution is consistent with both the BdG equations and the gap equation.

The total density and particle number are given by:
\begin{equation}
\rho(x) = \rho_\uparrow(x) + \rho_\downarrow(x), \qquad
N = \int dx \, \rho(x).
\end{equation}
In order to study pairing correlations in regions with $g=0$, we introduce the pair correlation function defined as
\begin{equation}
F(x) =  \langle \psi_\downarrow(x) \psi_\uparrow(x) \rangle.
\end{equation}
When \( g = g(x) \), we have
$\Delta(x) = -g(x) F(x)$. Consequently, 
$F(x) = \frac{1}{2} {\sum_n}' \left[ 
u_\uparrow^n(x) \, v_\downarrow^{n*}(x) 
+ u_\downarrow^n(x) \, v_\uparrow^{n*}(x) 
\right]$.

To normalize the physical quantities, we introduce a two-component non-interacting Fermi gas with particle numbers $N_\sigma=(N/2)$. The
density $\rho_0=N/(2L)$ gives the Fermi wavevector and energy of the non-interacting Fermi gas as
$k_F^0 = \pi \rho_0$ and  $E_F^0 = \frac{(\hbar k_F^0)^2}{2m}$.
The following dimensionless quantities defined:
$\tilde{g} = -g \frac{k_F^0}{E_F^0}$, 
$\tilde{h} = \frac{h}{E_F^0}$, and 
$\tilde{\Delta}(x) = \frac{\Delta(x)}{E_F^0}$. 
The pair correlation function $F(x)$ has units of inverse length. To render it dimensionless, we define
$\tilde{F}(x) = F(x) \cdot L$,
where $L$ is the box size.

\subsection{Fourier transform of the pair correlation function}
To analyze the momentum-space structures of pairing in the presence of population imbalance, we compute the discrete Fourier transform of $\tilde{F}(x)$ using a fast Fourier transform (FFT):
\begin{equation}
\tilde{F}(k_n) = \sum_j \tilde{F}(x_j) e^{-ik_n x_j} \Delta x,
\end{equation}
where $x_j$ are the spatial grid points and $\Delta x$ is the spatial grid spacing. Since $\tilde{F}(x)$ is dimensionless and the summation approximates an integral, the resulting $\tilde{F}(k)$ has units of length.
The wavevector is given by 
$k_n = \frac{2\pi n}{L}, \quad n = 0, 1, 2, \dots, \frac{N}{2}$. We emphasize that the Parseval theorem
$\sum_j |\tilde{F}(x_j)|^2 \Delta x = \sum_n |\tilde{F}(k_n)|^2 \Delta k$ with $\Delta k = 2\pi / L$ is satisfied by the including the $\Delta x$ in the FFT.
This convention allows us to compare the peaks in momentum space to physically meaningful momenta such as the Fermi wavevectors \( k_{F\uparrow} \) and \( k_{F\downarrow} \).
We remark that in inhomogeneous systems the densities \(\rho_\sigma(x)\) vary near
interfaces and hard-wall boundaries. To ensure that \(k_{F\sigma}\) remains a
meaningful reference scale, we define \(k_{F\sigma}\) from the bulk density plateau
on the relevant side of the system via \(k_{F\sigma}=\pi\rho_\sigma\).

Following the definition,  $|\tilde{F}(k)|^2$ represents the spectral weight in momentum space. To present dimensionless quantities, we rescale again using the non-interacting Fermi wavevector $k_F^0$ and define
$\tilde{F}(\tilde{k}) = k_F^0 \, \tilde{F}(k)$, $\tilde{k} = \frac{k}{k_F^0}$.
Due to the hard walls from the box potential, rapid oscillations near the edges of the box may arise due to sudden changes of the wave functions. Nevertheless, those transient behaviors vanish towards the bulk of the system. Therefore, we will present the Fourier transform of the bulk pair correlation function in the following discussions to focus on the signatures of coexistence of superfluid and normal phases in real space. Moreover, we have performed the corresponding numerical calculations using periodic boundary condition and confirmed our findings of the bulk remain the same in all cases studied here.

\subsection{Characteristic lengths and momenta}
In a population-imbalanced attractive Fermi gas, several characteristic scales govern the spatial structures when superfluid and normal phases encounter each other. These include length scales associated with pairing and momentum scales related to the Fermi surface mismatch.
While the BCS superfluid typically resists population imbalance, the spatial variation of the order parameter is characterized by the coherence length \cite{Fetter_book,deGennes_book,abrikosov1988}
\begin{equation}
\xi_{\text{BCS}} \sim \frac{\hbar^2 k_F}{m \Delta},
\end{equation}
where \( \Delta \) is the bulk gap function and $ k_F$  is the bulk Fermi wavevector. This sets the characteristic scale over which \( \Delta(x) \) varies. 
In the absence of population-imbalance, $\xi_{\text{BCS}}$ also determines the decay of $F(x)$ into the normal regime~\cite{Parajuli2023a}. However, we will see that the presence of the FFLO phase introduces a momentum scale.
The FFLO phase features spatially modulated pairing with a finite center-of-mass momentum. 
The dominant modulation appears at the wavevector~\cite{Yang2001,Rizzi2008}
$q = k_{F\uparrow} - k_{F\downarrow}$.
We introduce the dimensionless form
\begin{equation}
\tilde{q} = \frac{k_{F\uparrow} - k_{F\downarrow}}{k_F^0},
\end{equation}
which arises from pairing between the fermions on the mismatched Fermi surfaces~\cite{Fulde1964, Larkin1965}.  
This finite-momentum pairing is a signature of the FFLO state with more details given in Appendix~\ref{app:FFLO}, and we will show that it persists even in an inhomogeneous system.  
In contrast, the BCS phase features zero-momentum pairing and therefore shows no spectral peak at \(q\).

In order to distinguish the FFLO, BCS, and proximity-induced pair correlations in inhomogeneous systems, we will analyze  
key diagnostic quantities, including the coherence length \( \xi_{\mathrm{BCS}} \), the spin-resolved Fermi wavevectors \( k_{F\uparrow} \) and \( k_{F\downarrow} \), and the FFLO pairing momentum \( q = k_{F\uparrow} - k_{F\downarrow} \).  
Taken together, these quantities allow us to identify what kinds of pairing correlations can survive when inhomogeneous pairing interaction or spin polarization leads to complex structures involving different superfluid or normal phases coexisted in a box.


\begin{figure}[t]
    \includegraphics[width=\columnwidth]{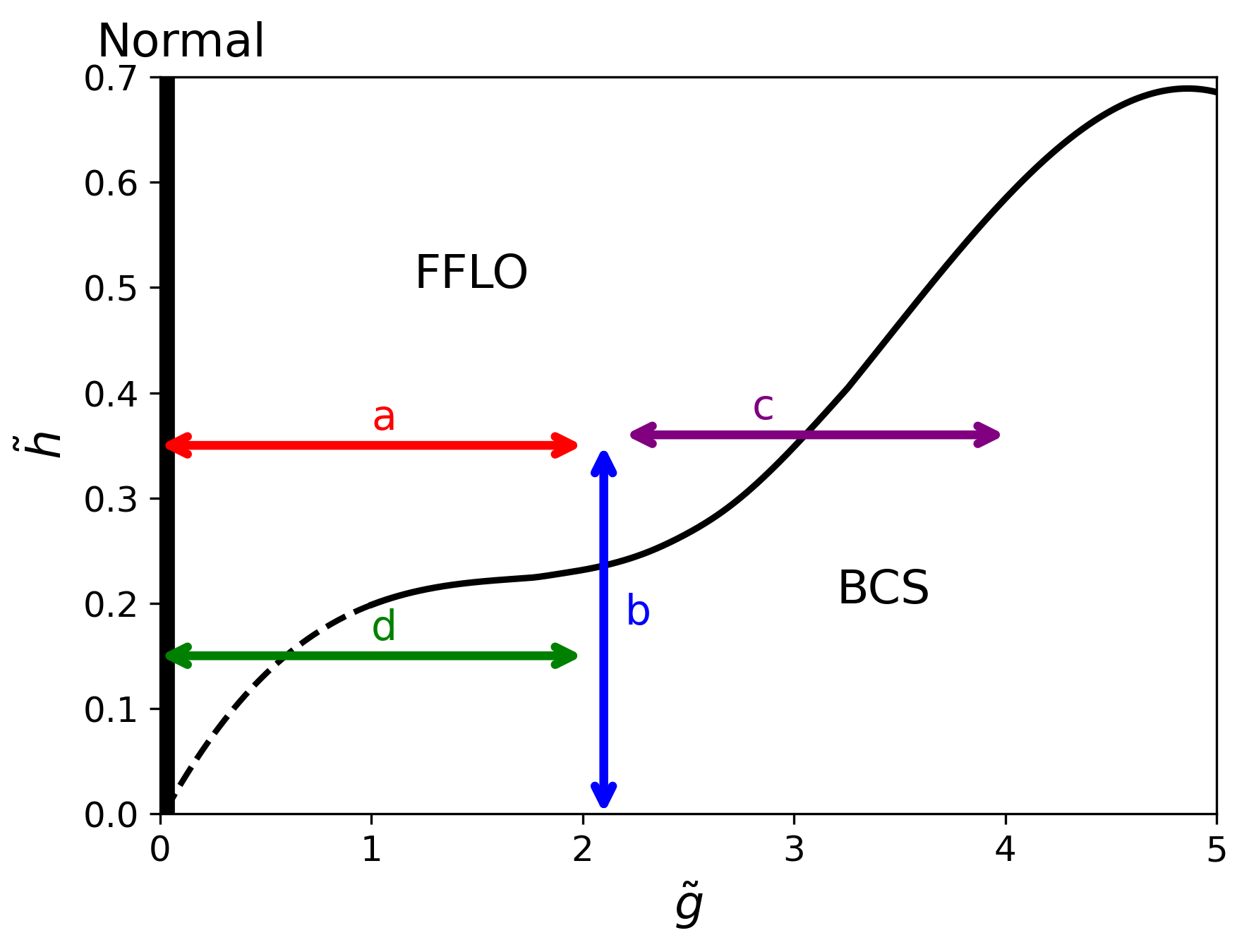}  
    \caption{
Phase diagram of two-component Fermi gases in a 1D box with uniform attractive coupling constant \( \tilde{g} \) and spin-polarization field \( \tilde{h} \). 
The BCS and FFLO superfluid phases as well as the normal phase are labeled. The arrows (a)–(d) indicate later constructions of inhomogeneous systems by joining two phases together using an inhomogeneous profile of \( \tilde{g} \) or \( \tilde{h} \). The dashed line in the lower-left corner represents our best estimations due to the blurring features when both pairing and spin-polarization are small.  
}
    \label{fig:phase_diagram}
\end{figure}

\section{Results}\label{Sec:Results}
\subsection{Phase Diagram}
We begin by presenting in Figure~\ref{fig:phase_diagram} the phase diagram of a two-component spin-imbalanced Fermi gas with uniform polarization and attractive interaction in a 1D box. In the absence of spin-polarization (\( \tilde{h} = 0 \)), the system is a BCS superfluid with a spatially uniform order parameter when \( \tilde{g}>0\) and an unpolarized normal Fermi gas when \( \tilde{g}=0\). 

As the spin-polarization \( \tilde{h} \) increases, population imbalance destabilizes the BCS phase and induces a transition to the FFLO state~\cite{Fulde1964,Larkin1965}. In this regime, Cooper pairs acquire finite center-of-mass momentum, and the order parameter becomes spatially modulated.  
The FFLO phase is particularly robust in 1D due to perfect nesting between the spin-resolved Fermi surfaces and enhanced phase space for finite-momentum pairing~\cite{Orso2007,Hu2007}. This contrasts with higher dimensions, where FFLO behavior is limited to a narrow sliver of parameter space. 
Meanwhile, a polarized normal phase still occupies the $\tilde{g}=0$ line.

We remark that the phase boundaries and structures of Fig.~\ref{fig:phase_diagram} are consistent with those from Refs.~\cite{Orso2007,PhysRevB.110.085159} in the corresponding intermediate-polarization range.
At higher \( \tilde{h} \), pairing can be fully suppressed, and the system transitions to a polarized normal state with non-vanishing $\tilde{g}$. However, we focus here on the intermediate-polarization regime and leave the high-polarization cases for future studies.

We now introduce inhomogeneity by joining two phases inside a box. This can be achieved by imposing different parameters on the two halves of the box. The colored arrows in Fig.~\ref{fig:phase_diagram} label the parameters joined together to explore inhomogeneous effects with interfaces and crossovers between the FFLO, BCS, and normal phases. Case (a) joins the FFLO and normal phases by introducing a real-space jump of the pairing strength in the presence of uniform spin polarization. Case (b) (case (c)) joins the FFLO and BCS phases by a real-space jump of spin polarization (pairing strength) with uniform pairing strength (spin polarization). Case (d) joins the BCS and normal phases by a real-space jump of the pairing strength with relatively low, uniform spin polarization. 

In the following, we will analyze the four selected cases with different joint structures and present the results from the simulations with $N_x=300$ grid points since we have checked that further increasing $N_x$ does not qualitatively change our conclusions. Moreover, slight changes of $\tilde{\mu}$ and $\tilde{h}$ shift the densities, but all qualitative features are found to remain the same.

\subsection{Proximity effect between FFLO and normal phases}
\begin{figure}[t]
    \centering
    \includegraphics[width=\columnwidth]{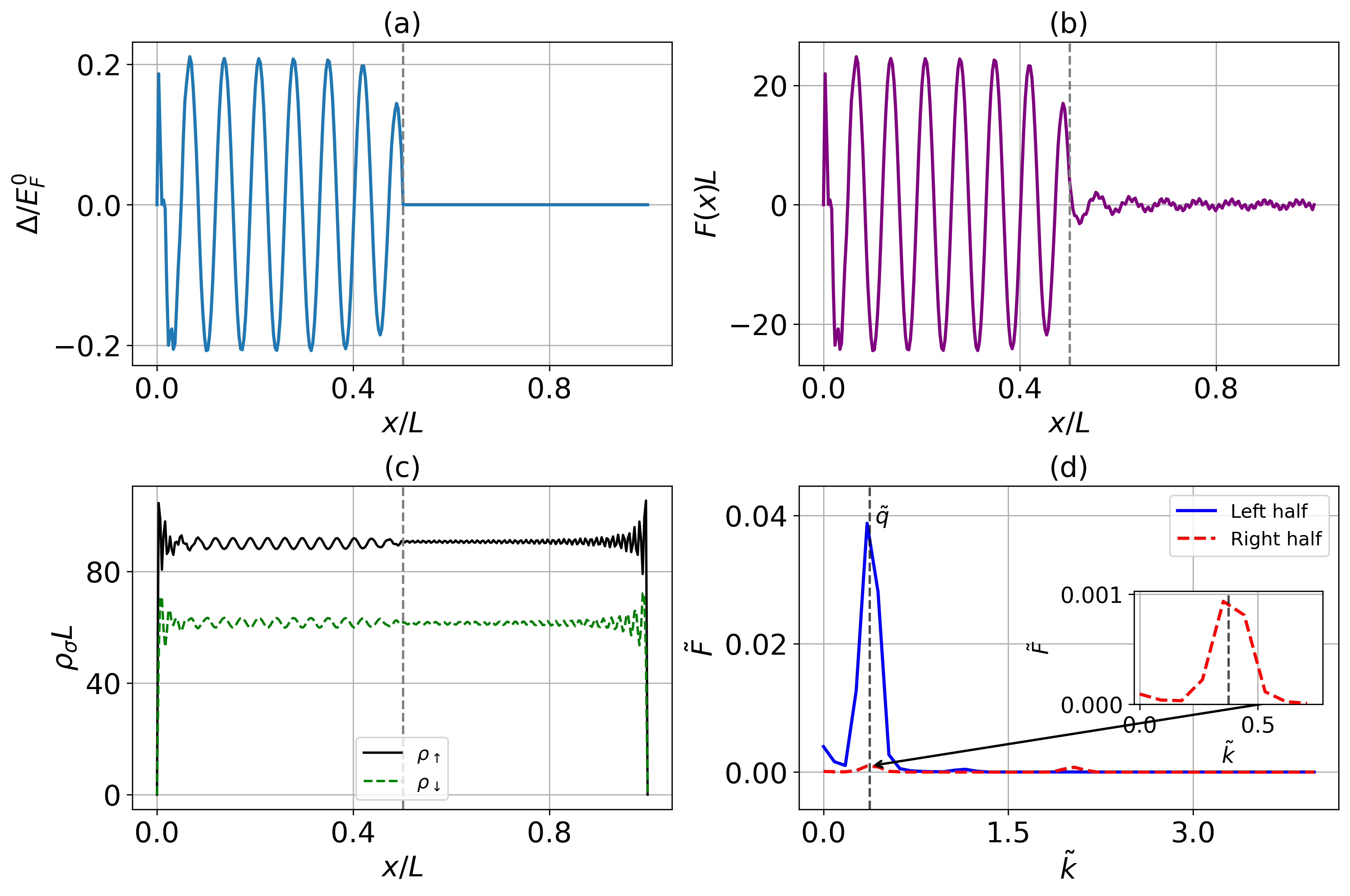}
    \caption{Joint FFLO–normal configuration generated by an inhomogeneous pairing 
interaction profile with $\tilde{g}_L = 2$ and $\tilde{g}_R = 0$, 
as indicated by arrow (a) in Fig.~\ref{fig:phase_diagram}. 
The vertical dashed lines mark where $\tilde{g}$ changes. 
Here $\tilde{\mu}=1$ and $\tilde{h}=0.35$ are uniform. (a) The order parameter $\Delta(x)$ displays the FFLO oscillations 
on the left half and vanishes in the normal region. (b) The pair correlation function $F(x)$ exhibits oscillations on the left side and penetrates into the noninteracting region.(c) Spin-resolved densities $\rho_\sigma(x)$ reflect the population imbalance.
(d) Dimensionless Fourier amplitudes $\tilde{F}(\tilde{k})$ of the left 
(blue solid line) and right (red dash line) halves of $F(x)$.  
The FFLO region shows a sharp peak at the FFLO momentum 
$\tilde{q} = (k_{F\uparrow} - k_{F\downarrow})/k_F^0$.  
The normal region also has a weaker peak at $\tilde{q}$ (magnified in the inset),
demonstrating proximity-induced FFLO correlations.  }\label{fig:FFLO_normal_transition}
\end{figure}

We first consider the structure joining an FFLO phase and a noninteracting normal phase, which is created by a step-function profile of the pairing interaction \( \tilde{g}(x) \). 
In the left half of the system with \( \tilde{g} = 2 \), the ground state features FFLO pairing whereas the right half with \( \tilde{g} = 0 \) remains normal. 
Figure~\ref{fig:FFLO_normal_transition} summarizes the spatial profiles of the order parameter \( \Delta(x) \), pair correlation function \( F(x) \), spin-resolved densities \( \rho_\uparrow(x) \), \( \rho_\downarrow(x) \), and the Fourier transform of \( F(x) \).

The order parameter in Fig.~\ref{fig:FFLO_normal_transition}(a) shows periodic oscillations on the FFLO side and vanishes in the normal region due to the absence of pairing interactions. 
Nevertheless, as shown in Fig.~\ref{fig:FFLO_normal_transition}(b), the pair correlation function does not vanish abruptly at the interface. 
Instead, it penetrates into the noninteracting region while retaining the oscillatory structure from the FFLO phase. 
This behavior reflects the superconducting proximity effect~\cite{deGennes1969ProximityEffects,Tinkham}, but here we generalize to a population-imbalanced system in which the pairing correlations carry the finite momentum characteristic of the FFLO phase.

The momentum-space analysis in Fig.~\ref{fig:FFLO_normal_transition}(d) highlights the penetration of the FFLO pairing correlation into the normal region. 
The Fourier transform of \( F(x) \) on the FFLO side (blue solid curve) exhibits a sharp peak at $\tilde q$,
which is the hallmark of finite-momentum pairing due to the mismatch between the spin-polarized Fermi surfaces.  
Remarkably, the result in the normal region (red dashed curve), despite having no intrinsic pairing interaction, also shows a weaker but discernible peak at the same \( \tilde q \).  
This demonstrates that the proximity-induced pairing correlations not only leak into the normal region but also preserve the finite-momentum character due to population imbalance, in contrast to the conventional BCS proximity effect where the pair correlations in the normal regime show smooth decay without modulations~\cite{Parajuli2023a}.


The spin-resolved densities shown in Fig.~\ref{fig:FFLO_normal_transition}(c) remain partially polarized on both sides of the interface.  
The transition from the FFLO region to the normal region does not produce sharp signatures in the densities, since the amplitude of the FFLO order parameter in Fig.~\ref{fig:FFLO_normal_transition}(a) is relatively small and therefore does not strongly modify the distributions of the fermions.

Taken together, these results illustrate that the FFLO pair correlation can propagate across the interface and imprint its oscillatory character onto the normal region.  
The presence or absence of the spectral peak at the FFLO momentum \( \tilde q \) in the Fourier transform of \( F(x) \) provides a clear and robust diagnostic for distinguishing the FFLO finite-momentum pairing from the uniform BCS pairing and establishes the proximity effect between the FFLO and normal phases due to population imbalance.

\subsection{Proximity effect between FFLO and BCS phases}
Next, we analyze the structures when the BCS and FFLO superfluid phases are joined together. There are multiple ways to achieve such a mixture, and we discuss two orthogonal protocols separately in the following. Those setups allow us to isolate the effects of spin imbalance and pairing to study how the two distinct superfluid phases interact across their interface.

\subsubsection{Inhomogeneity by \( \tilde{h} \)}
\begin{figure}[t]
    \centering
    \includegraphics[width=\columnwidth]{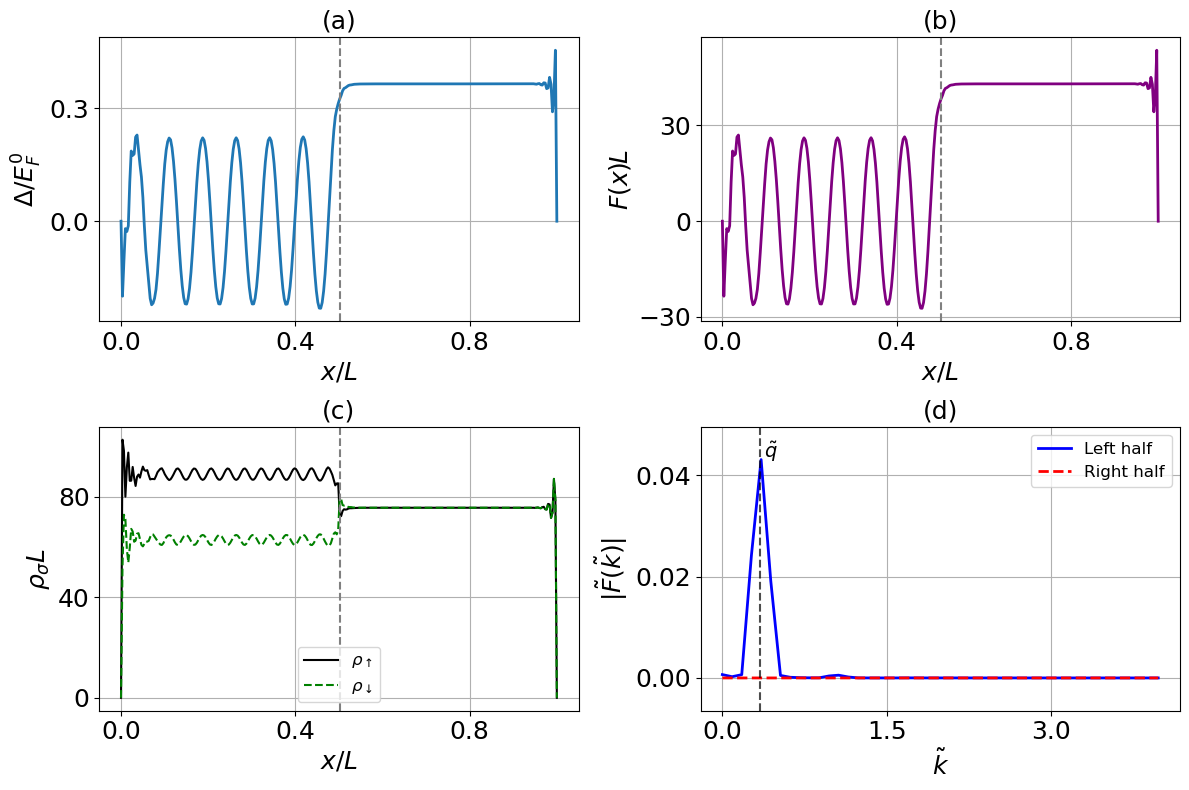}
   \caption{
    Joint FFLO–BCS configuration generated by an inhomogeneous profile of $\tilde{h}_L=0.35, \tilde{h}_R=0$ for the left and right halves of the system, corresponding to arrow $b$ in Fig.~\ref{fig:phase_diagram}. Here $\tilde{\mu} = 1$ and $\tilde{g} = 2$ are uniform. The vertical dashed lines mark where $\tilde{h}$ changes.
    (a) The order parameter $\Delta(x)$ exhibits oscillations (a plateau) in the FFLO (BCS) region. 
    (b) The pair correlation function $F(x)$ shows similar behavior. 
    (c) The spin-resolved densities $\rho_{\sigma}(x)$ are imbalanced (balanced) on the FFLO (BCS) side.
    (d) Fourier spectra of the bulk $F(x)$ on the left (solid) and right (dashed) sides.
The left FFLO region shows a sharp peak at the FFLO momentum $\tilde q$ while the right BCS region exhibits no observable feature.}

    \label{fig:FFLO_BCS_interface_h}
\end{figure}

We first introduce a spatially varying spin polarization field \( \tilde{h}(x) \), such that the left (right) half supports the FFLO (BCS) phase with a fixed uniform pairing strength. 
For such a case, Figure~\ref{fig:FFLO_BCS_interface_h}(a) shows that the order parameter \( \Delta(x) \) exhibits modulations on the FFLO side but remains uniform in the BCS region. 
In the case shown in Fig.~\ref{fig:FFLO_BCS_interface_h}, the spin polarization field vanishes on the BCS side, thereby excluding population imbalance by construction. 
Meanwhile, the pair correlation function \( F(x) \), shown in Fig.~\ref{fig:FFLO_BCS_interface_h}(b) transitions smoothly from the FFLO oscillations to the uniform BCS plateau, with no sharp discontinuity at the interface. This reflects the coherence of the Cooper pairs and a continuity of the order parameter across the interface.

In contrast, Fig.~\ref{fig:FFLO_BCS_interface_h}(c) shows the spin-resolved densities, which behave quite differently across the BCS--FFLO interface. The density difference \( \rho_\uparrow(x) - \rho_\downarrow(x) \) is suppressed in the BCS region due to the vanishing spin-polarization field but becomes visible in the FFLO region since the oscillations of the order parameter can result in oscillatory density distributions and population imbalance. Interestingly, population imbalance spills over into a tiny sliver on the BCS side, barely visible on the plot.

The Fourier transforms of the bulk pair correlation functions on the two sides, shown in Fig.~\ref{fig:FFLO_BCS_interface_h}(d), reveal the spectral distinction between the two types of Fermi superfluids. 
On the FFLO side (blue solid line), the dominant peak appears at the FFLO wavevector \(\tilde q\), reflecting the finite-momentum pairing driven by the mismatched Fermi surfaces.  
In contrast, the BCS side (red dashed line) shows no appreciable spectral weight at any finite momentum, consistent with the zero-momentum Cooper pairs whose bulk Fourier spectrum is essentially featureless.

Since the FFLO parameter oscillates with the length scale $1/\tilde{q}$ while the BCS coherence length is 
\( \xi_{\mathrm{BCS}} \), we compare those length scales to confirm that the FFLO pairing correlation does not penetrate deeply into the BCS region.  
We
evaluate the dimensionless BCS coherence length using the width of the $5\%\!-\!95\%$ range of the order parameter on the BCS side near the BCS-FFLO interface and obtain
$ \xi_{\mathrm{BCS}}^{(5\text{–}95)} k_F^0 
\simeq 2.37$.  
For comparison, the FFLO momentum extracted from the FFLO side gives $1/\tilde{q} \simeq 2.91$.  
Thus, $q\xi_{\mathrm{BCS}}^{(5\text{–}95)}$ is less than $1$.
Therefore, the order parameter resumes the zero-momentum BCS form before any FFLO oscillations set in near the interface, which explains why the bulk BCS pair correlations function in Fig.~\ref{fig:FFLO_BCS_interface_h}(d) does not exhibit the FFLO momentum peak.


\subsubsection{Inhomogeneity by \( \tilde{g} \)}
Another way to join the FFLO and BCS phases in real space is to introduce a spatially varying profile of the pairing strength \( \tilde{g}(x) \) while holding the spin polarization field \( \tilde{h} \) fixed. For a selected case with \( \tilde{g}(x) = 2 \) on the left half (FFLO side) and \( \tilde{g}(x) = 4 \) on the right half (BCS side) with a constant \( \tilde{h}=0.35 \), the setup probes how pairing evolves across a transition where the local pairing interaction increases across the interface under the same spin polarization field.

\begin{figure}[t]
    \centering
    \includegraphics[width=\columnwidth]{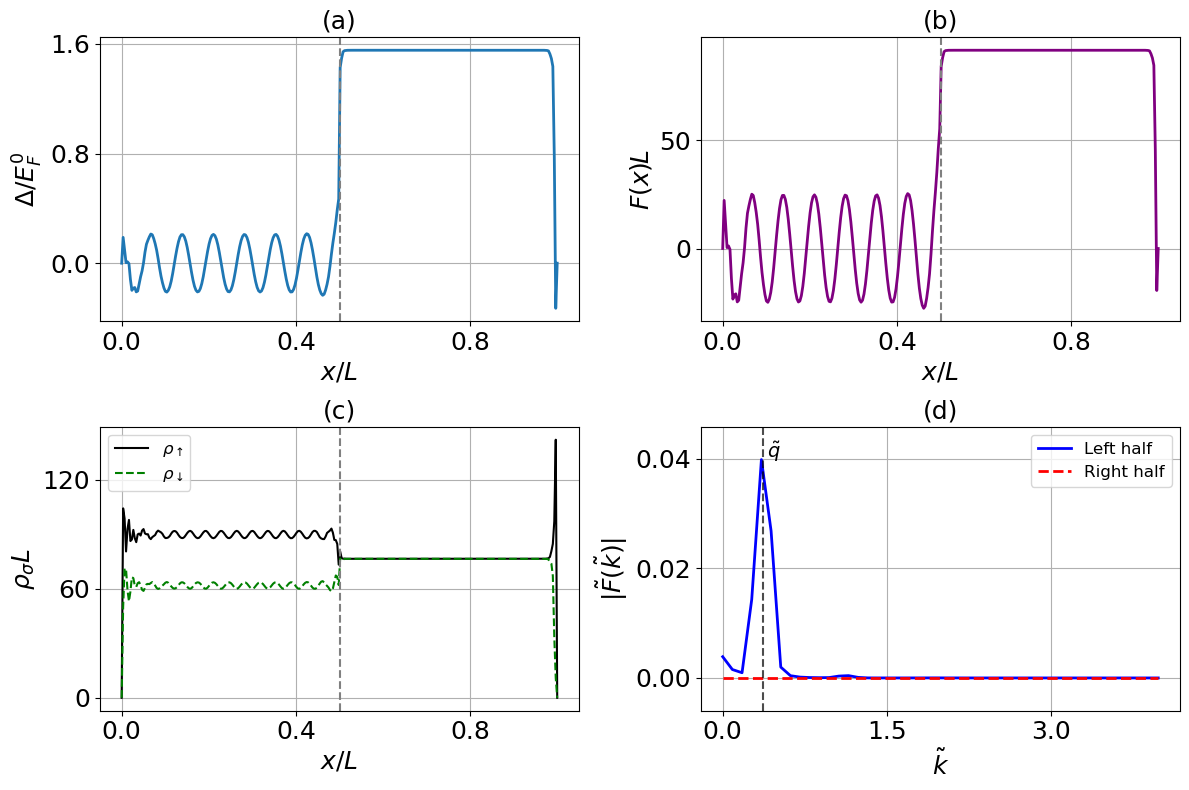}
    \caption{
    Joint FFLO–BCS configuration by an inhomogeneous profile of $\tilde{g}_L=2, \tilde{g}_R=4$ for the left and right halves of the system, corresponding to arrow $c$ in Fig.~\ref{fig:phase_diagram}. The vertical dashed lines mark where $\tilde{g}$ changes. Here $\tilde{\mu} = 1$ and $\tilde{h} = 0.35$ are uniform. 
    (a) The order parameter $\Delta(x)$ exhibits oscillations (a plateau) in the FFLO region (BCS) region. 
    (b) The pair correlation function $F(x)$ follows a similar structure. 
    (c) The densities $\rho_{\sigma}(x)$ are imbalanced (balanced) in the FFLO (BCS) region with fluctuations near the hard-wall boundaries. 
    (d) Fourier spectra of the bulk $F(x)$ on the left (solid line) and right (dashed line) sides, showing the FFLO momentum $\tilde{q}$ on the left and a flat line on the right.
}
    \label{fig:FFLO-BCS-interface_g}
\end{figure}

As shown in Fig.~\ref{fig:FFLO-BCS-interface_g}(a), the order parameter \( \Delta(x) \) exhibits spatial oscillations in the FFLO region and becomes uniform on the BCS side. Unlike the previous case with an inhomogeneous spin polarization field, here the enhancement of \( \Delta(x) \) on the BCS side is due to the increase of the pairing strength across the interface. 
The corresponding pair correlation function \( F(x) \), shown in Fig.~\ref{fig:FFLO-BCS-interface_g}(b), demonstrates a continuous crossover from the FFLO finite-momentum pairing to the BCS zero-momentum pairing, even in the presence of a uniform spin polarization field. 

In the momentum-space spectrum shown in Fig.~\ref{fig:FFLO-BCS-interface_g}(d), the Fourier transform of \( F(x) \) in the bulk confirms the dominant momenta of the two superfluids. The FFLO side shows a peak at the FFLO momentum \(\tilde q \) while the BCS side exhibits a strong zero-momentum peak. The suppression of the FFLO peak on the BCS side demonstrates that even in the presence of a moderate spin polarization field, sufficiently strong pairing interactions can maintain the BCS phase, consistent with previous studies of homogeneous systems in 1D~\cite{Orso2007, Hu2007}. To quantify the FFLO-BCS interface in this case, we compare the BCS coherence length
\(\xi_{\mathrm{BCS}}\) on the BCS side with the characteristic FFLO modulation length
\(1/q\) extracted from the dominant finite-momentum peak in Fig.~\ref{fig:FFLO-BCS-interface_g}(d) on the FFLO side. 
The case shown in Fig.~\ref{fig:FFLO-BCS-interface_g} yields \(\xi_{\mathrm{BCS}}\, q \approx 0.48\) from the bulk gap and \(\xi_{\mathrm{BCS}}^{(5\text{-}95)}\, q \approx 0.29\) from substituting $\xi_{BCS}$ by the width of the \(5\%\)–\(95\%\) range of \(F(x)\). When \(\xi_{\mathrm{BCS}}\, q < 1\), the BCS condensate is stiff on the scale of the FFLO
modulation, so the oscillatory pairing does not penetrate deeply into the BCS region. 
Therefore, a spatial quench of the pairing strength with uniform spin polarization can generate a relatively sharp FFLO–BCS interface with distinct features on each side.


Due to the finite spin polarization field throughout the system and the suppression of the order parameter near the hard-wall boundaries of the box, tiny polarized normal regions develop near both ends of the box. These regions are characterized by a local depletion of the gap function and the appearance of population imbalance, most clearly visible in the spin-resolved densities near the hard walls shown in Fig.~\ref{fig:FFLO-BCS-interface_g}(c), which resembles Fig.~\ref{fig:FFLO_BCS_interface_h}(c). Such boundary effects in a box already emerge in homogeneous systems, as summarized in Appendix~\ref{App:BoundaryEffects}.
However, small regions of population imbalance appear within the BCS region, both near the interface with the FFLO phase and close to the hard-wall boundary when the order parameter vanishes, as shown in Fig.~\ref{fig:FFLO-BCS-interface_g}(a) while the spin polarization field is present.

\subsection{Proximity effect between BCS and normal phases}
One can see from Fig.~\ref{fig:phase_diagram} that in the presence of a finite spin-polarization field, the BCS and normal phases are always separated by the FFLO phase in a homogeneous setting.
We now analyze the intriguing case where the BCS phase is joined by a spin-imbalanced normal phase by a spatial quench of the pairing strength \( \tilde{g}(x) \) under a fixed but relatively small spin-polarization field \( \tilde{h} \). Consequently, the left half of the box supports an unpolarized BCS phase while the right half is occupied by a polarized normal gas. This corresponds to the combination labeled \( d \) in Fig.~\ref{fig:phase_diagram}. Here we address the question on joining two phases not adjacent on the homogeneous phase diagram by using the BCS and polarized normal phases as an example.

\begin{figure}[t]
    \centering
    \includegraphics[width=\columnwidth]{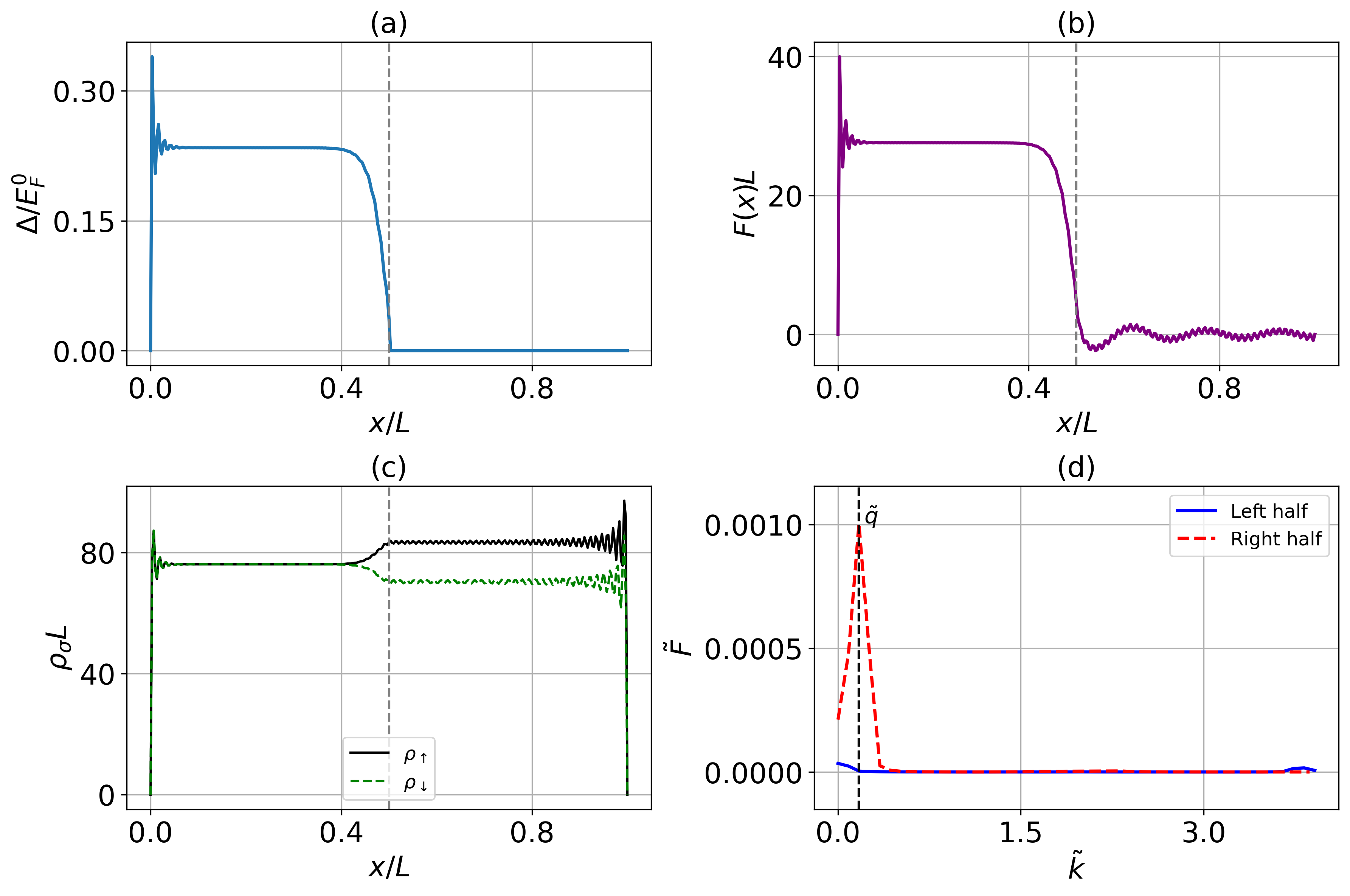}
    \caption{Joint BCS-normal configuration generated by an inhomogeneous pairing-interaction profile 
    with $\tilde g_{L} = 2$ and $\tilde g_{R} = 0$ for the left and right halves, 
    corresponding to arrow $d$ in Fig.~\ref{fig:phase_diagram}. The vertical dashed lines mark where $\tilde{g}$ changes. 
    Here $\tilde \mu = 1$ and $\tilde h = 0.15$ are uniform. (a) The order parameter $\Delta(x)$ is nearly uniform in the 
    BCS region and vanishes in the normal region. (b) The pair correlation function $F(x)$ shows a plateau on the BCS side but exhibits oscillations characteristic of the FFLO pair correlation.  
    (c) The spin-resolved densities $\rho_{\sigma}(x)$ are balanced in the BCS region and 
    become imbalanced in the normal region. (d) Bulk Fourier spectra of $F(x)$ from the left (solid line) and right (dashed line) halves.  
    The BCS side is featureless but the normal side displays a peak at the FFLO momentum
    $\tilde{q}$, indicating a buffer of the FFLO phase as the system changes from the BCS to polarized normal regions.
}\label{fig:BCS_normal_interface}

\end{figure}

As shown in Fig.~\ref{fig:BCS_normal_interface}(a), the order parameter \( \Delta(x) \) is uniform and finite in the BCS region but drops rapidly toward zero across the interface due to the vanishing pairing interaction on the normal side. Nevertheless, Fig.~\ref{fig:BCS_normal_interface}(b) shows that the pair correlation function \( F(x) \) penetrates into the normal region, where it decays gradually but exhibiting spatial oscillations indicating the FFLO pair correlation.

We again analyze the interfacial properties by comparing the BCS coherent length and the FFLO momentum. From Fig.~\ref{fig:BCS_normal_interface}(a), we extract 
\( \xi_{\mathrm{BCS}}^{(95\rightarrow \mathrm{if})} \) in the BCS region from the distance between the point where \( \Delta(x) \) reaches \(0.95\,\Delta_{\mathrm{bulk}} \) and the interface position \(x_{\mathrm{if}}\) while the FFLO momentum $q$ is estimated from the spin-polarized densities in the bulk of the normal region. 
In this case, we obtain
\( \xi_{\mathrm{BCS}} q \simeq 2.66 \).
Thus, the BCS order parameter decays over a couple of the FFLO modulations, thereby allowing the finite-momentum pair correlations to be generated and transmitted through the interface rather than being suppressed abruptly in the previous BCS-FFLO cases.

The spin-resolved densities in Fig.~\ref{fig:BCS_normal_interface}(c) further elucidate the nature of the transition region. Although a finite spin-polarization field is present throughout the system, the population imbalance remains sufficiently weak on the BCS side, maintaining nearly equal spin densities \( \rho_\uparrow(x) \approx \rho_\downarrow(x) \). Therefore, the BCS side is resistant to population imbalance by the conventional singlet pairing. In contrast, the normal-gas side exhibits finite population imbalance due to the spin polarization field.

Given the abrupt drop of the pairing interaction that nominally allows only the BCS and polarized normal phases, the oscillatory features of \( F(x) \) in Fig.~\ref{fig:BCS_normal_interface}(b) are not remnants of a preexisting FFLO phase. Instead, they are generated locally at the transition region since the suppression of the BCS order parameter gives rise to a buffer zone supporting the FFLO behavior in between the BCS and normal phases.
The momentum-space signature of this induced FFLO pairing is shown in Fig.~\ref{fig:BCS_normal_interface}(d). In the bulk BCS region, \( |F(k)| \) exhibits only a zero-momentum peak, whereas on the normal side a clear finite-momentum peak appears at the FFLO momentum \( \tilde q \), reflecting the mismatch of the spin-dependent Fermi surfaces and confirming the proximity-induced FFLO correlations even though there is no bulk FFLO phase in the setting.
Consequently, despite the effort to join the BCS and normal phases in the presence of spin-polarization by a spatial quench of the pairing interaction, the dropping BCS order parameter at the interface allows a buffer region with the FFLO pairing correlations. Furthermore, the FFLO pair correlation penetrates the normal region and exhibits proximity effect. Therefore, a BCS-FFLO-normal structure emerges as a compromise to the continuous changes of physical quantities in real space.

\section{Implications}\label{Sec:Implications}

For Fermi gases in one dimension, exact solutions of the Gaudin--Yang model
\cite{Gaudin1967,Yang1967} show partially polarized ground states with
power-law decaying pairing correlations and possible phase diagrams
\cite{Orso2007,Hu2007,GuanBatchelorLee2013}.
Bosonization and Luttinger-liquid approaches further elucidate the nature of
correlations and the absence of genuine long-range order in one dimension
\cite{Yang2001,Giamarchi2004}.
Numerical studies, including the density-matrix renormalization group 
calculations \cite{Feiguin2007,Rizzi2008,TezukaUeda2008} and quantum Monte Carlo
simulations \cite{Batrouni2008}, have provided complementary insights into FFLO
correlations, pairing structures, and phase separation in lattice and continuum
models.
Furthermore, 1D Fermi gases have pronounced features coming from the point-like
spin-dependent Fermi surfaces, which strongly constrain the available pairing
phase space and render the system highly sensitive to local population imbalance
and interaction strength, as emphasized in previous studies of homogeneous
systems~\cite{Orso2007,Hu2007}.
Nevertheless, quasi-1D Fermi gases have been shown to support superfluid or other
phases in their ground states~\cite{Liao2010,Revelle2016}, and the inhomogeneous
systems studied here introduce rich physics to spin-polarized Fermi gases.

Experimentally, spatial control of interactions and effective spin-polarization
fields in ultracold atoms may be achieved using optical and magnetic techniques.
Optical Feshbach resonance enables local and dynamical tuning of the pairing
interaction~\cite{Fatemi2000,Theis2004}.
More recently, optically controlled magnetic Feshbach resonance schemes have
provided improved stability and flexibility for spatial and temporal control
of interactions~\cite{Bauer2009,Clark2015,Jagannathan2016,Arunkumar2019}.
Alternatively, magnetic-field gradients provide a complementary route with
longer coherence times~\cite{DiCarli2020}.

The momentum-resolved pair correlations have been shown useful for identifying
proximity effects from the BCS or FFLO phases in inhomogeneous Fermi gases with
population imbalance.
These correlations may be accessible using momentum-resolved radio-frequency
(rf) spectroscopy~\cite{Stewart2008,Gupta2003,Shin2007} to infer the pairing gap
function.
In addition, density-density correlations may provide an alternative route to
the detection of pairing correlations of cold-atom systems~\cite{Altman2004,
Greiner2005}.

To confirm that the qualitative features of the inhomogeneous structures and
interfacial properties of spin-polarized Fermi gases are robust against the
details of the spatial profiles of the pairing interaction or spin polarization,
we have performed additional numerical calculations with more gradual spatial
changes of the parameters, such as a linear ramp of the pairing interaction or
spin polarization with a finite transition region.
The resulting order parameter, pair correlation functions, and momentum-space
peaks are qualitatively the same as the corresponding cases obtained with the
step-function changes of the parameters if the transition widths of the
parameters do not exceed the BCS coherence length or the inverse of the FFLO
momentum.
Therefore, the emergence and penetration of various pair correlations discussed
here are genuine proximity effects rather than an artifact from the idealized,
abrupt drop of the parameters.

The robustness of the results against the parameter profiles also makes the
results presented here testable in experiments, where the parameters are
switched on or off within a finite range of distance.
For example, optical or magnetic Feshbach resonance~\cite{Bauer2009,Fatemi2000,
Theis2004,Clark2015,Jagannathan2016,Arunkumar2019} can tune the interaction, while
single-site or spatially resolved spin manipulations demonstrated by spin
rotations~\cite{Weitenberg2011}, spin-selective transfer~\cite{Gupta2003}, and
spin flips (see Ref.~\cite{Kuhr2016} for a review) may be adapted to realize
inhomogeneous spin polarization for studying proximity effects.
Therefore, population-imbalanced atomic Fermi gases provide a flexible and
controllable platform for exploring proximity-induced phenomena and the
interplay between Fermi superfluidity and magnetism beyond conventional
electronic systems.

\section{Conclusion}\label{Sec:Conclusion}
We have presented exemplary spatial structures when inhomogeneous pairing interaction or spin polarization is applied to a quasi-1D two-component Fermi gas in a box potential by using the BdG equation to take into account superfluidity and population imbalance. The coexistence of the FFLO, BCS, and normal phases in real space results in different interfacial and bulk behaviors as the parameters change. The intrinsic momentum and length scales, including the spin-resolved Fermi wave vectors, FFLO momentum, and BCS coherence lengths, determine the structures around the interfaces and the penetrations of pair correlations in different settings. While the FFLO momentum in the pair correlation determined by population imbalance is visible in the FFLO-normal and BCS-normal structures, the BCS phase typically remains unpolarized and uniform in the bulk. The emergence of a buffer FFLO phase between the BCS and normal phases in the presence of population imbalance further illustrates the rich physics induced by inhomogeneity. Our results elucidate various proximity effects realizable in highly controllable cold-atom systems with measurable features and shed light on multi-phase structures in real space.

To inspire future exploration, the central idea of introducing inhomogeneous structures in
multi-component atomic systems can be adopted and generalized to investigate many relevant
phenomena, such as atomic Bose--Bose~\cite{PhysRevLett.81.1543,PhysRevA.59.1457,PhysRevA.94.013602,PhysRevA.100.063623} or Bose--Fermi~\cite{PhysRevA.66.013614,FerrierBarbut2014,DeSalvo2019,Parajuli2023} mixtures.
This line of research is expected to unveil and characterize complex interplay between many-body
physics, inhomogeneity, and phase transitions.

\begin{acknowledgments}
D.J.G. and B.P. were partially supported by the William and Linda Frost Fund in the Cal Poly Bailey College of Science and Mathematics.
C.C.C. was supported by the NSF (Grant No. PHY-2310656) for the theoretical analysis and the DOE (Grant No. DE-SC0025809) for the physical implications. B.P. used ANVIL at Rosen Center for Advanced Computing (RCAC) at Purdue University through allocation PHY250194 from the Advanced Cyberinfrastructure Coordination Ecosystem: Services and Support (ACCESS) program, which is supported by U.S. National Science Foundation grants No. 2138259, 2138286, 2138307, 2137603, and 2138296.
\end{acknowledgments}

\appendix
\section{FFLO phase with uniform parameters}\label{app:FFLO}
\begin{figure}[t]
    \centering
    \includegraphics[width=\columnwidth]{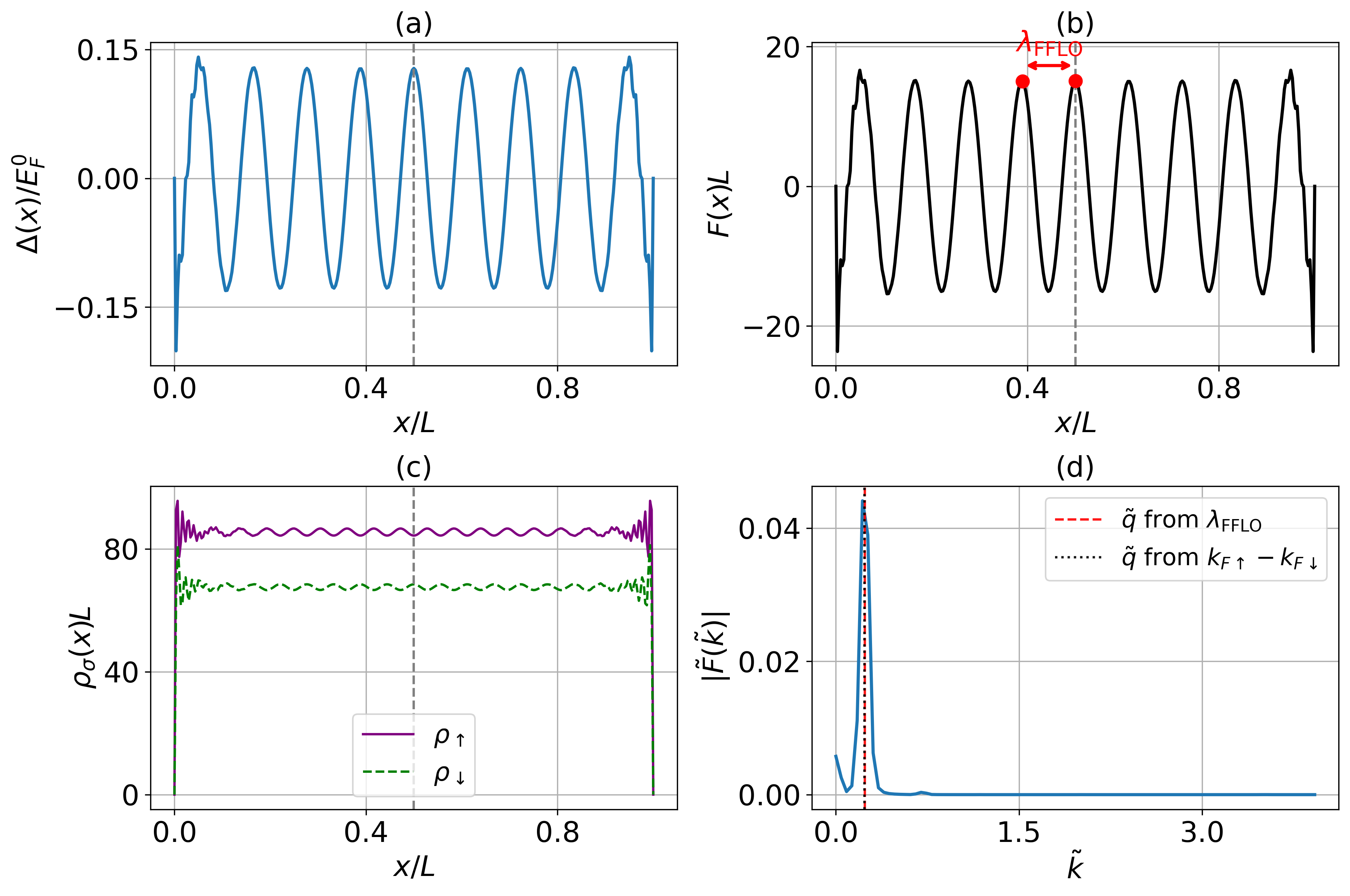}
    \caption{
        FFLO phase in a 1D box induced by uniform polarization field $\tilde h=0.24$, $\tilde \mu=1$, and pairing strength $\tilde g=2$.
        The order parameter $\Delta(x)$ (a) and the pair correlation function $F(x)$ (b) show the signature oscillations. The wavelength determine by adjacent peaks gives  $\lambda_{\mathrm{FFLO}}$.
        (c) The spin-resolved densities $\rho_{\sigma}(x)$ exhibits population imbalance due to $\tilde h$.
        (d) Fourier transform of $F(x)$ shows a peak at the FFLO momentum $\tilde q$.}
    \label{fig:fflo_homo}
\end{figure}

Figure~\ref{fig:fflo_homo} presents an exemplary FFLO-phase solution of the BdG equation for a two-component Fermi gas with uniform pairing strength \( \tilde{g} = 2 \), chemical potential \( \tilde{\mu} = 1 \), and spin-polarization field \( \tilde{h} = 0.24 \) in a 1D box. 
Figure~\ref{fig:fflo_homo}(a) shows the signature oscillatory order parameter \( \Delta(x) \). This is the hallmark of FFLO pairing due to population imbalance. 
Meanwhile, the pair correlation function \( F(x) \) shown in Fig.~\ref{fig:fflo_homo}(b) exhibits the same oscillatory structure because of the uniform $\tilde g$. The peak-to-peak distance of \( F(x) \) can be used to extract the FFLO wavelength \( \lambda_{\mathrm{FFLO}} \), which implies a wavevector \( q^{\text{real}} = 2\pi/\lambda_{\mathrm{FFLO}} \). This provides a comparison with the theoretical FFLO wavevector \( q^{\text{theory}} = k_{F\uparrow} - k_{F\downarrow} \) derived from the spin-resolved Fermi momenta \( k_{F\sigma} = \pi \rho_\sigma \).
We find in  Fig.~\ref{fig:fflo_homo} that $\tilde q_{\mathrm{real}}\simeq 0.238$ and
$\tilde q_{\mathrm{theory}}\simeq 0.234$, showing a good agreement.

The spin-resolved densities \( \rho_\sigma(x) \) plotted in Fig.~\ref{fig:fflo_homo}(c) show the presence of population imbalance due to the spin-polarization field. Notably, the majority species (spin-up fermions in this case) tends to accumulate at the nodes of the order parameter \( \Delta(x) \), where the energy gap vanishes. The spatial modulations of the densities give another characteristic feature of the FFLO state~\cite{Yang2001, Orso2007}.

Finally, the momentum-space spectrum shown in Fig.~\ref{fig:fflo_homo}(d) exhibits a pronounced peak in the Fourier transform of the pair correlation function \( F(x) \) at the finite wave vector \( q = k_{F\uparrow} - k_{F\downarrow} \). This dominant peak directly reflects the formation of finite-momentum FFLO pairing. The sharpness of the spectral feature indicates well-defined pairing correlations at a single dominant momentum, consistent with the periodic oscillations in both order parameter \( \Delta(x) \) and pair correlation function in real space.

Taken together, the real-space and momentum-space signatures demonstrate that, for homogeneous interactions and finite spin imbalance, the FFLO phase emerges naturally as the stable self-consistent solution of the BdG equations in a 1D box. The FFLO phase is characterized by an oscillatory order parameter and densities, along with a distinct FFLO momentum set by the mismatch of the Fermi wave vectors.

\begin{figure}[t]
    \centering
    \includegraphics[width=\columnwidth]{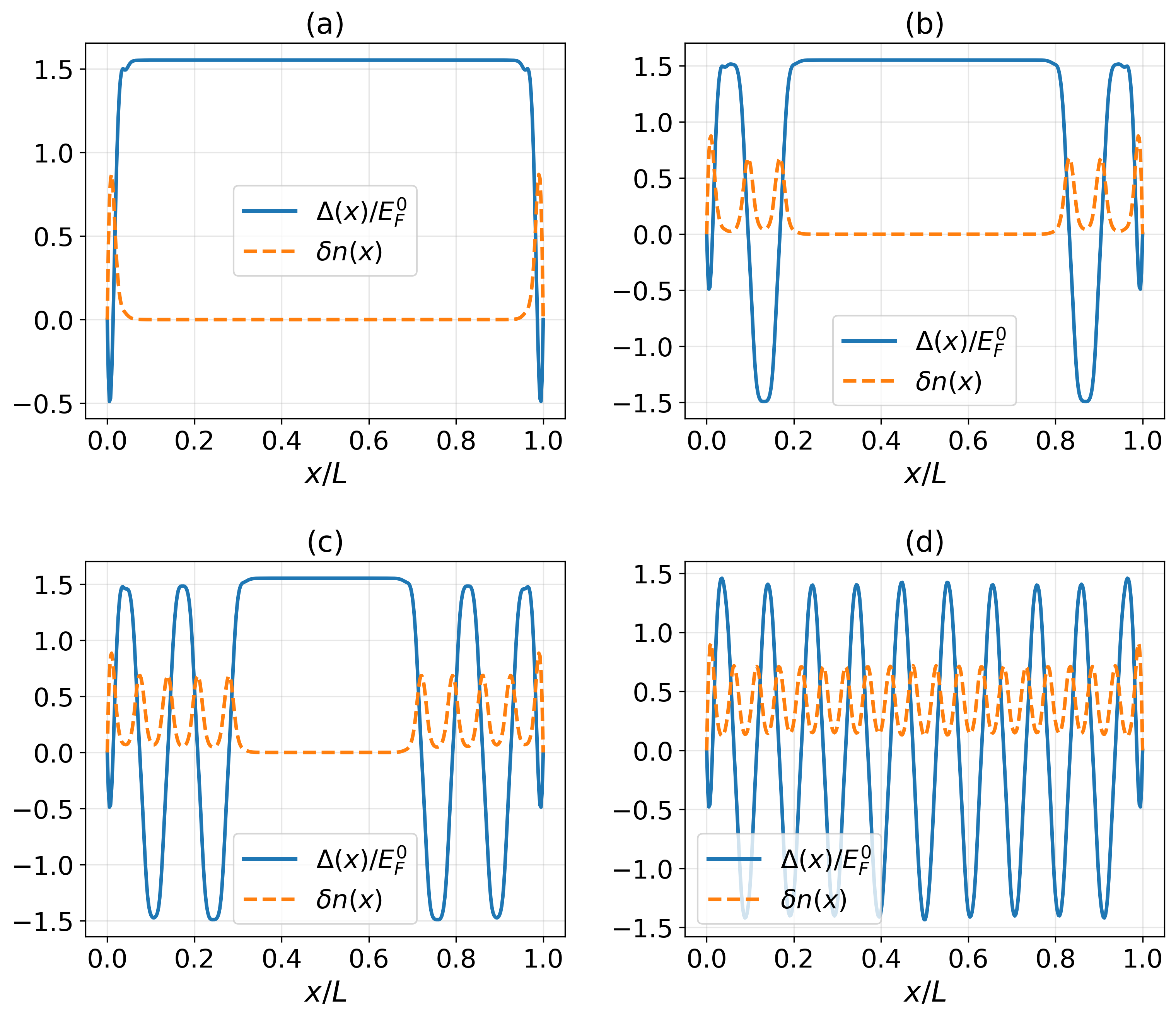}
   \caption{Pairing field $\Delta(x)$ for homogeneous 1D Fermi gases with $\tilde{g} = 5$, $\tilde{\mu} = 1$, and varying spin polarization fields $\tilde{h} = 0.1 (a), 0.5 (b), 0.6 (c), 0.7 (d)$. Although the parameters are uniform, the hard-wall boundaries induce spatial modulations in $\Delta(x)$ near the edges. With increasing population imbalance, FFLO oscillations become more pronounced and occupy from the edges to the bulk.}\label{fig:edge-effects}
\end{figure}

\section{Boundary effects in a box}\label{App:BoundaryEffects}
The hard-wall boundaries of a finite box impose constraints on the spatial structures of the Fermi gas inside since the order parameter, densities, and pairing correlations must vanish there. As a result, even in systems with homogeneous parameters, boundary effects may still arise near the hard walls.
To examine the boundary effects due to the hard walls, we consider a two-component Fermi gas confined in a 1D box with uniform parameters given by the chemical potential \( \tilde{\mu} = 1 \), interaction strength \( \tilde{g} = 5 \), and several values of the spin-polarization field \( \tilde{h} = 0.1,\,0.5,\,0.6,\) and \(0.7\). Figure~\ref{fig:edge-effects} displays the corresponding spatial profiles of the order parameter \( \Delta(x) \) and the density difference \( \delta n(x) = \rho_\uparrow(x) - \rho_\downarrow(x) \).

At low spin-polarization (\( \tilde{h} = 0.1 \)) as shown in Figure~\ref{fig:edge-effects}(a), the bulk away from the hard walls is predominantly in the BCS phase, characterized by a nearly uniform plateau of \( \Delta(x) \) and vanishing population imbalance. A tiny sliver of population-imbalanced region, however, appears near each hard wall since the vanishing order parameter there no longer expel spin polarization. As the spin-polarization field is increased, the system gradually evolves towards the FFLO phase, starting near the hard walls and moving towards the center. As shown in panels (b) and (c), the FFLO regions characterized by the oscillatory order parameter and density difference occupy more space from the hard walls while the bulk BCS region with a flat order parameter shrinks.

When the spin polarization is strong as shown in Figure~\ref{fig:edge-effects}(d), the FFLO modulations of the order parameter and density difference occupy basically the whole box. However, sharp distortions of the order parameter and density difference are still visible near the hard walls due to the rapid change of the wave functions.

The distortions near the hard walls of box potentials are specific to the geometry, as there is no such distortions in similar systems with periodic boundary conditions. We have performed simulations using both open and periodic boundary conditions to verify the bulk properties are the same. In the main text, we present the results in box potentials and focus on the bulk results in the Fourier transform to avoid distractions from the hard-wall boundary effects.

%

\end{document}